\documentclass[pdflatex,sn-mathphys-num]{sn-jnl}

\usepackage{mathrsfs}
\usepackage{graphicx}%
\usepackage{multirow}%
\usepackage{amsmath,amssymb,amsfonts}%
\usepackage{amsthm}%
\usepackage{mathrsfs}%
\usepackage[title]{appendix}%
\usepackage{xcolor}%
\usepackage{textcomp}%
\usepackage{manyfoot}%
\usepackage{booktabs}%
\usepackage{algorithm}%
\usepackage{algorithmicx}%
\usepackage{algpseudocode}%
\usepackage{listings}%

\theoremstyle{thmstyleone}%
\theoremstyle{thmstyletwo}%

\theoremstyle{thmstylethree}%

\begin{document}

\title[Article Title]{Topological Invariants in the Pore Morphology Method}

\author{\fnm{Fernando} \sur{Alonso-Marroquin}}

\email{fernando.marroquin@kfupm.edu.sa}

\affil{\orgdiv{CIPR}, \orgname{King Fahd University of Petroleum and Minerals}, \city{Dhahran}, \postcode{31261}, \country{Kingdom of Saudi Arabia}}


\abstract{
This study introduces a pore morphology algorithm that emphasizes the central role of topology in multiphase flow through porous media. Analysis of drainage in lattice-based pore networks identifies two key quantities, the percolation threshold and residual saturation, as topological invariants. These descriptors, which are based solely on connectivity rather than geometric details, capture the essential structure of the network. The percolation threshold is interpreted as a topological phase transition, marking the transition from global connectivity of the defending fluid to isolated clusters of trapped fluid. The universality of scaling exponents across different lattice geometries reveals the existence of topological universality classes, where systems with equivalent connectivity display identical critical behavior. This topological framework underscores the robustness of the identified invariants and provides a general basis for upscaling pore-scale processes in complex media.
}



\maketitle

\section{Introduction}\label{sec1}

The displacement of one fluid by another in a porous medium is a process that appears in fields as diverse as physics \cite{furuberg1988dynamics}, geology \cite{hunt2017flow}, and hydrology \cite{assouline1998conceptual}. It also plays a role in our daily routines. Every morning, the flavor of our coffee depends on whether hot water can percolate through packed grains and how effectively it extracts the rich compounds within. The first result is controlled by the percolation threshold, whether there is a connected pathway for flow, while the second is governed by residual saturation, the fraction of air pockets that remains trapped.
This study shows that, under pressure-controlled conditions, both of these key quantities are not determined by material properties or pore geometry, but rather by the topology of the network itself. They are topological invariants, unchanged under continuous deformations of the pore structure, and can be derived directly from percolation theory, where connectivity, rather than geometry, defines the essence of the process. Here,the term {\it topological invariant} is used to refer to quantities that depend solely on the connectivity structure of the pore network and remain unchanged under continuous deformations of geometry or change of material parameters. In this sense, these invariants belong to the broad class of phenomena described by percolation.

Percolation is one of the most documented emerging phenomena in complex systems. It appears when the nodes or bonds in a random network are progressively activated, creating connected clusters where complex patterns emerge \cite{furuberg1988dynamics}. In bond percolation theory, the control parameter $p$ is defined as the probability of bond occupancy in a network. A phase transition occurs at a critical value $p_c$. Above this value, large-scale connectivity begins to emerge in the lattice.
Perhaps the most exciting theoretical advance on bond percolation is the Kensen theorem \cite{kesten1980critical}
that provides a rigorous demonstration that the bond percolation threshold for infinite-square lattices is $p_c =1/2$. Kersten theorem confirmed previous analytical derivations by Sykes and Essan, who also showed bond percolation threshold of $2\sin{(\pi/18)}$, $1/2$, $1-2\sin{(\pi/18)}$ and for triangular, rectangular, and honeycomb lattices \cite{sykes1963some}. 

Wilkinson and Willemsen \cite{wilkinson1983invasion} introduced a variant of the percolation process known as invasion percolation (IP). In IP, invasion occurs at a constant flow rate, with the interface advancing through the path of least resistance at each step. When the defending fluid is incompressible, regions of it may become isolated and cannot be invaded further. This phenomenon, referred to as invasion percolation with trapping (IPT), is characterized by the invading fluid forming a single connected cluster that grows along the path of minimal resistance \cite{dias1986percolation}. Importantly, however, the IP and IPT processes are not topologically invariant: the invasion path depends explicitly on the geometry and material properties of the pore network, rather than solely on its connectivity.
These results align well with real flow-rate-controlled situations, where residual saturation is influenced by factors such as wettability and capillary number \cite{humphry2014impact,wolf2020influence,nguyen2006effect}. In contrast, in the quasistatic limit controlled by pressure, it is unclear how residual saturation after primary drainage is related to the material and geometrical properties of the porous medium.
To address this gap, a pore morphology method is proposed based on a topological framework, where capillary pressure saturation is related to topological invariants, pore throat distribution, and interfacial properties of solids and liquids. The residual saturation and the percolation threshold emerge as invariants that depend solely on network connectivity, regardless of geometry or material properties.

The pore morphology method (PMM), introduced by Hilpert and Miller \cite{hilpert2001pore}, provides a different model in which fluid invasion is calculated using morphological operations on binary images. More recently, an extended version of PMM was proposed to account for trapping \cite{alonso2025capillary}. This model simulates the displacement of an incompressible defending fluid by an invading phase. The invasion process exhibits critical features that resemble both the ordinary percolation transition \cite{stauffer2018introduction} and the backbone percolation transition \cite{sampaio2018elastic}. As in IPT, the invading fluid forms a single cluster connected to the injection zone. However, unlike IPT, the invasion does not follow the path of least resistance; instead, it fills the available pore space under pressure control, leaving clusters of defending fluid disconnected from the outlet and therefore trapped. Thus, while IP describes flow-rate-controlled invasion, PMM captures the pressure-controlled quasistatic regime.

To date, the relationship between the percolation transition and residual saturation in PMM and the underlying topology of porous media has not been established. Nor is it clear to what extent these features are governed by material parameters of the pores and fluids. Establishing whether the transition corresponds to ordinary percolation, backbone percolation, or a distinct class of topological transition remains an important step toward applying PMM as a predictive tool for multiphase flow.

This paper is organized as follows. Section \ref{sec:topology} formulates the pore morphology method as a topological percolation model. Section \ref{sec:results} presents the calculation of the pressure-saturation relation in two-dimensional lattices. Section \ref{sec:conclusion} highlights the topological invariants with universal scaling exponents identified by this new pore morphology approach and discusses how these enable general laws for multiphase flow that are independent of geometry and material properties.

\section{Topological Percolation Model}
\label{sec:topology}
As in invasion percolation (IP), it is assumed that the porous medium is initially filled with a resident fluid and is progressively invaded by another fluid in an immiscible and incompressible fashion \cite{knackstedt2002nonuniversality}. Different from IP, the invasion is controlled by pressure, not flowrate. The porous medium is modeled as a network of spherical pores connected by cylindrical throats. Blunt \cite{blunt2017multiphase} presented the connection between fluid invasion during drainage and bond percolation (BP). There, the quasi-static displacement of the liquid in a throat of radius $r$ is resolved by the equilibrium between capillary forces $P_c \times\pi r^2$ and surface tension $\gamma\cos\theta \times 2\pi r$,
where $P_c$ is the capillary pressure, $\gamma$ and $\theta$ are the interfacial tension and the contact angle. Thus, the capillary pressure required to invade the throat is $P_c = 2\gamma\cos\theta/r$. For a given capillary pressure, the {\it potentially active} throats are defined as these whose radius satisfies $r>r_c$, where $r_c$ is the so-called {\it capillary radius}: 
\begin{equation}
    r_c=\frac{2\gamma\cos\theta}{P_c}
    \label{eq:capillary radius}
\end{equation}
To establish the link between quasi-static fluid invasion and BP, the undirected graph of the pore network $g$ is defined as the collection of all pores (nodes) and throats (edges). In the same way, we define the {\it allowed graph} from the subset $g_0 \subseteq g$ of all potentially active throats, that is, the subset of the graph whose edge radii satisfy $r>r_c$. The sub-graph $g_0$ defines the pore space that can potentially be invaded, which is different from the pore space that is actually occupied \cite{heiba1992percolation}. Whether a pore in $g_0$ can be invaded depends on the connectivity between the invading and defending fluids and the injection/evacuation pores. For the purpose of simulating invasion, Blunt \cite{blunt2017multiphase} defines occupancy as the fraction of throats that belong to $g_0$, that is, $p = N_0/N_t$, where $N_0$ is the number of active throats and $N_t$ is the total number of throats. based on Eq.~\ref{eq:capillary radius}, the occupation probability can be calculated in terms of the cumulative distribution function of the throat radii $F(r)$ as:
\begin{equation}
    p = 1-F(r_c).
    \label{eq:occupancy}
\end{equation}
 Eq.~\ref{eq:occupancy} leads to a similarity between quasi-static invasion and BP, where percolation is controlled by the occupation probability $p$. A percolation threshold $p_c$ is expected, above which a phase transition occurs, leading to the emergence of a giant connected component in the pore network. The order parameter in bond percolation is typically defined as the probability that a given node belongs to the largest cluster~\cite{kesten1980critical}. In our case, it is more convenient to use another order parameter that is the saturation;  This is defined as the fraction of pores that are invaded, $S_I = N_{p,I}/N_p$.  Since we simulate drainage, all the nodes connected to the invaded throats are invaded. Note that this saturation is an upper bound of the actual saturation of the wetting fluid $S_w$. This is because when the wetting fluid reaches a pore, it may not be fully occupied but a fraction of the pore space may remain occupied by the resident fluid due to the capillary and trapping effects \cite{blunt2017multiphase}.

The fluid invasion algorithm presented here is an improved version of the Alonso-Marroquin \& Andersson  pore morphology method (PMM) ~\cite{alonso2025capillary}. The PMM calculates the invasion based on morphological operations and connectivity functions on the binary image of the porous medium. Here, an improvement of the method is proposed by formulating the problem on the undirected graph instead.

The fundamental graph operators are defined as follows. The erosion of a graph, $g_0 = g \ominus S_r$, removes all throats (edges) with radii smaller than $r$, leaving only those pores (nodes) that can be accessed by a capillary of radius $r$. The connectivity function, $g_c = C(g_1,g_2)$, returns the subset of $g_1$ that is connected to $g_2$, thus identifying the accessible clusters from a specified boundary or reservoir. The computational complexity of these operations is $O(zN_p\log{N_p})$, where $z = 2N_t/N_p$ is the average coordination number, $N_t$ is the number of throats, and $N_p$ is the number of pores.

To account for residual saturation, the pore space is divided into three disjoint subgraphs: invaded $g_I$, defended $g_D$, and trapped $g_T$, such that $g = g_I \cup g_D \cup g_T$. During drainage, pores that lose connection to the evacuation zone are classified as trapped. Boundary conditions are imposed by assigning certain nodes to the reservoir (R), where the invading fluid is injected, and others to the evacuation zone (E), where the resident fluid exits. Initially, the entire pore space belongs to the defended graph ($g_D = g$), with the occupancy probability set to zero.

Each step of the quasistatic simulation is implemented as follows:

\begin{enumerate}[noitemsep]
\item Increment the occupancy $p$ and determine the corresponding capillary radius $r_c$ from Eq.\ref{eq:capillary radius}.
\item Construct the allowed graph as $g_0 = g \ominus S_{r_c}$, retaining only throats larger than $r_c$, and expand it to include adjacent nodes.
\item Update the invaded region by adding the part of $g_0$ that is not trapped and remains connected to the reservoir: $g^I_{new} = C(g_0 - g_T, R)$; then set $g_I \leftarrow g_I \cup g^I_{new}$.
\item Recalculate the defended region as the portion of the graph that is not invaded and remains connected to the evacuation zone: $g_D = C(g - g_I, E)$.
\item Identify newly trapped pores as those that belong neither to the invaded nor defended region: $g^T_{new} = g - (g_D \cup g_I)$; then update $g_T \leftarrow g_T \cup g^T_{new}$.
\item Repeat the procedure from Step 1 until the target occupancy $p=1$ is achieved.
\end{enumerate}

This algorithm differs from IP algorithms that use the shortest-path algorithm \cite{sheppard1999invasion, wilkinson1986percolation} in Step 3. In flow-rate controlled situations, the invaded fluid may choose the shortest path of the allowed pore space. This is different from the pressure-controlled situation, where all allowed pores connected to the invaded zone are invaded in each step. This condition has been validated using the PMM \cite{hilpert2001pore}.  The computational complexity of the shortest-path IP algorithm is $O(N_t\log{N_t})$~\cite{sheppard1999invasion}, while the algorithm proposed here is $O(zN_cN_p\log N_p)$. Here, $N_c$ is the number of pressure increments. As in IP, the proposed PMM algorithm presents sleek advantages over other methods for multiphase flow in porous media, such as the lattice Boltzmann method \cite{connington2012review}, the level-set computational fluid dynamics \cite{olsson2005conservative}, and classical pore morphology methods \cite{hilpert2001pore}. This is because the method is quasistatic, mesh-free, pixel-free, and has an unprecedented reduction of the computational complexity since the operations are performed in the extremely reduced dataset-- the graph of the pore network. 

\section{Results}
\label{sec:results}

\begin{figure}[t]
    \centering
    \includegraphics[trim={6cm 10cm 6cm 8cm},clip,width=0.45\linewidth]{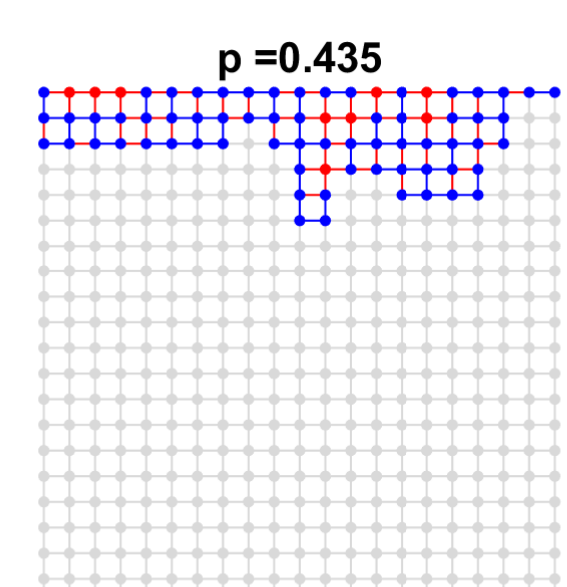}
    \includegraphics[trim={6cm 10cm 6cm 8cm},clip,width=0.45\linewidth]{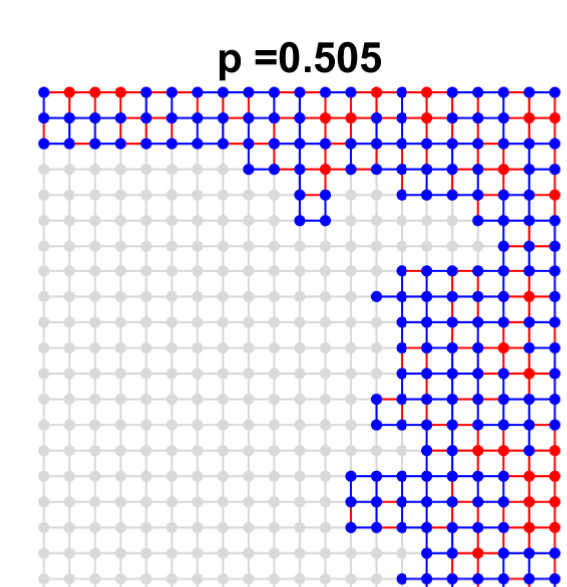}
      \includegraphics[trim={6cm 10cm 6cm 8cm},clip,width=0.45\linewidth]{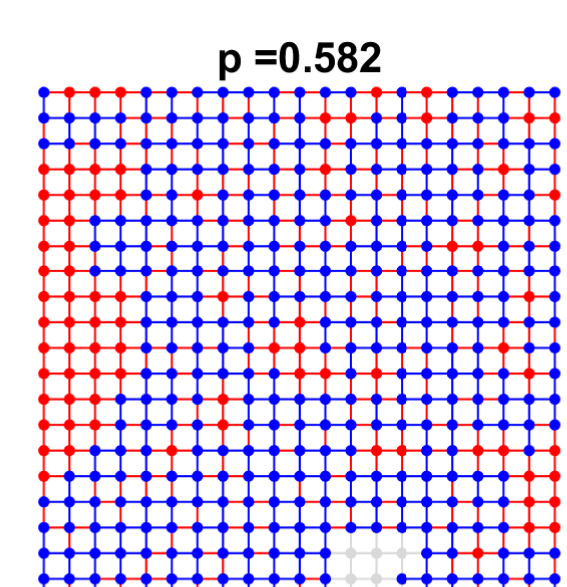}
       \includegraphics[trim={6cm 10cm 6cm 8cm},clip,width=0.45\linewidth]{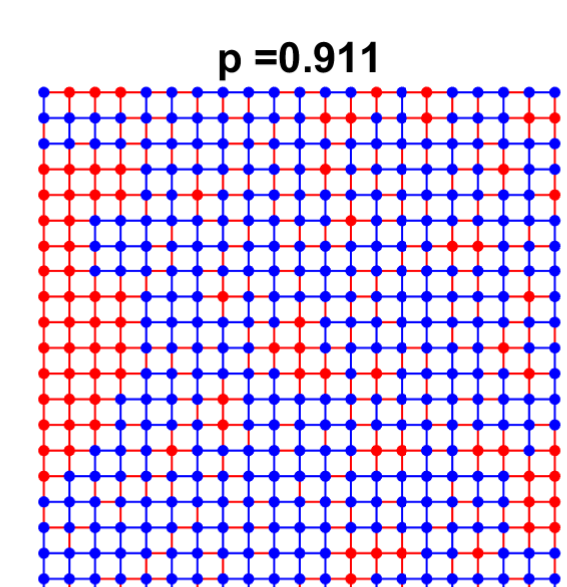}
    \caption{Snapshots of the pressure-controlled lattice invasion on a square lattice. The defended, invaded, and trapped phases are represented by gray, blue, and red colors, respectively. The first snapshot corresponds to the subcritical regime, $p < p_c$. The second snapshot is taken around the critical threshold, $p \approx p_c$. The third snapshot captures the moment when the invaded network (backbone) is fully established, and the last snapshot shows the final stage of drainage. The top/bottom nodes are connected to the injection/evacuation zone.}
    \label{fig:snapshots}
\end{figure}

Some snapshots of a simulation with a square lattice are shown in Fig.~\ref{fig:snapshots}. The upper nodes are connected to the injection zone and the lower one with the evacuation zone. The first stage of the simulation corresponds to the subcritical regime ($p<p_c$), which is characterized by slow growth of the invaded region with minimal trapping. The percolation transition occurs near $p_c$ and marks the point where the percolation occurs. After percolation, a short regime is detected, starting with the growth of the initial invaded percolation cluster and ending when the invaded cluster is fully established, creating a sub-network where the flow of invaded fluid is fully established. This is often referred to as the backbone \cite{sampaio2018elastic}. The final stage is characterized by a slow filling of the invaded and trapped fluid that ends with a residual saturation, that is, the final saturation of the resident fluid: $S_r = 1-S^{final}_I>0$. Note that both critical percolation $p_c$ and residual saturation $S_r$ are calculated directly from the graph, making them topological invariants. They depend only on the topology of the network and are independent of its geometry or the material properties of the solid matrix and fluids.

Similarly to invasion percolation (IP), the invading fluid in the PMM originates as a small cluster and grows as the probability of occupation increases. However, unlike IP, this cluster does not advance along a narrow path of least resistance but instead fills all connected pore spaces linked to the reservoir. Trapping occurs when the defending fluid is isolated in disconnected clusters that lose connectivity and remain immobile, surrounded by the invading phase. In contrast, in bond percolation (BP), clusters emerge randomly as the bonds are occupied. Multiple disconnected clusters co-exist, and at the percolation threshold a single cluster abruptly spans the system, while the others remain finite. In this statistical framework, one may loosely refer to invaded clusters (occupied bonds) and defended clusters (unoccupied bonds), but these are defined probabilistically rather than dynamically. No genuine trapping occurs; both types of clusters simply coexist as complementary subsets of the lattice, with the spanning cluster signifying the transition. This comparison highlights the distinct nature of the processes: PMM is governed by hysteretic (history-dependent) trapping; IP is characterized by dynamic growth with trapping; and BP represents a purely statistical occupation process, devoid of both dynamic evolution and hysteretic trapping.

\subsection{Finite size effects}

Finite-size effects are intrinsic to all percolation processes \cite{hunt2017flow,stauffer2018introduction,sampaio2018elastic}. In IP and PMM, the system size constrains the growth of the invading cluster and the statistics of trapped regions, while in BP finite-size scaling controls cluster-size distributions and the likelihood of forming a spanning cluster. These effects are especially relevant for multiscale flow in porous media, since laboratory and numerical studies are always limited to finite systems. Understanding how finite-size scaling links pore-scale invasion to macroscopic transport laws is therefore essential for reliable upscaling.

\begin{figure}[t]
    \centering
    \includegraphics[trim={4cm 7cm 3cm 5cm},clip,width=0.32\linewidth]{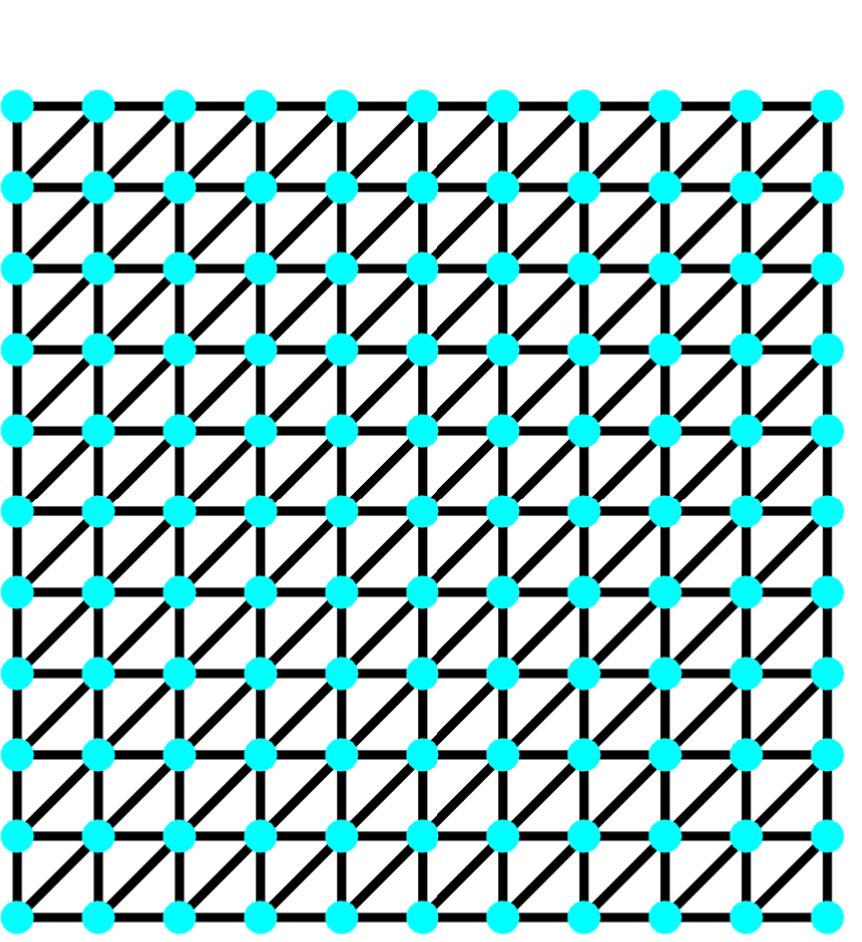}
    \includegraphics[trim={4cm 7cm 3cm 5cm},clip,width=0.32\linewidth]{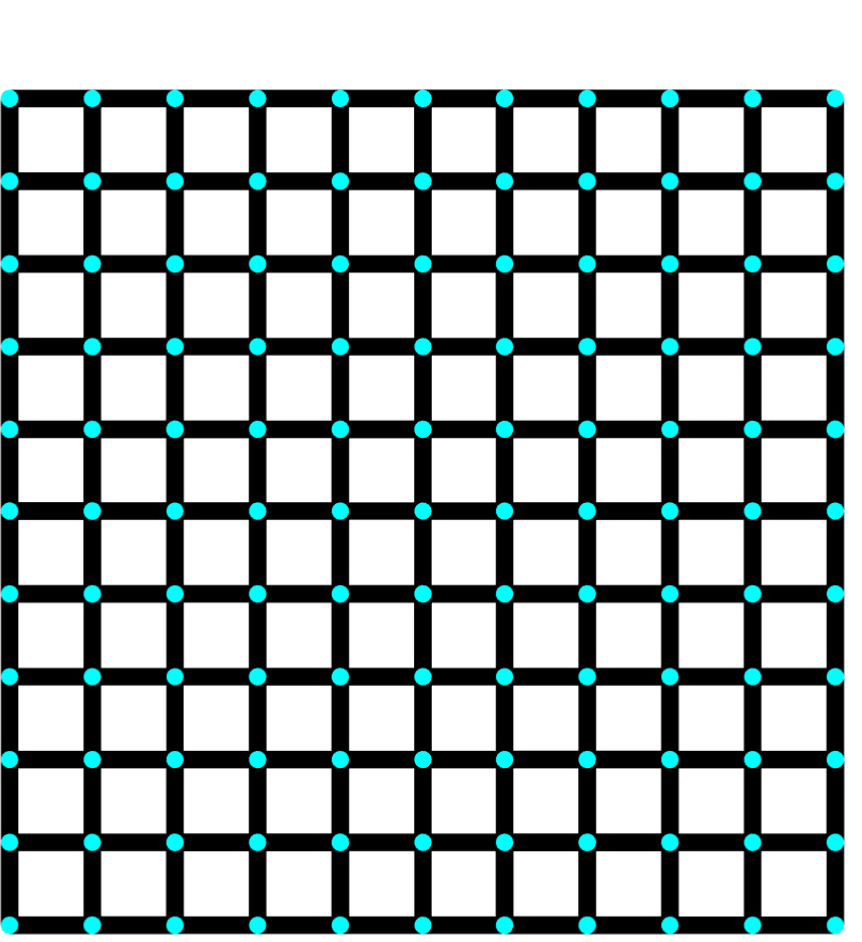}
    \includegraphics[trim={4cm 7cm 3cm 5cm},clip,width=0.32\linewidth]{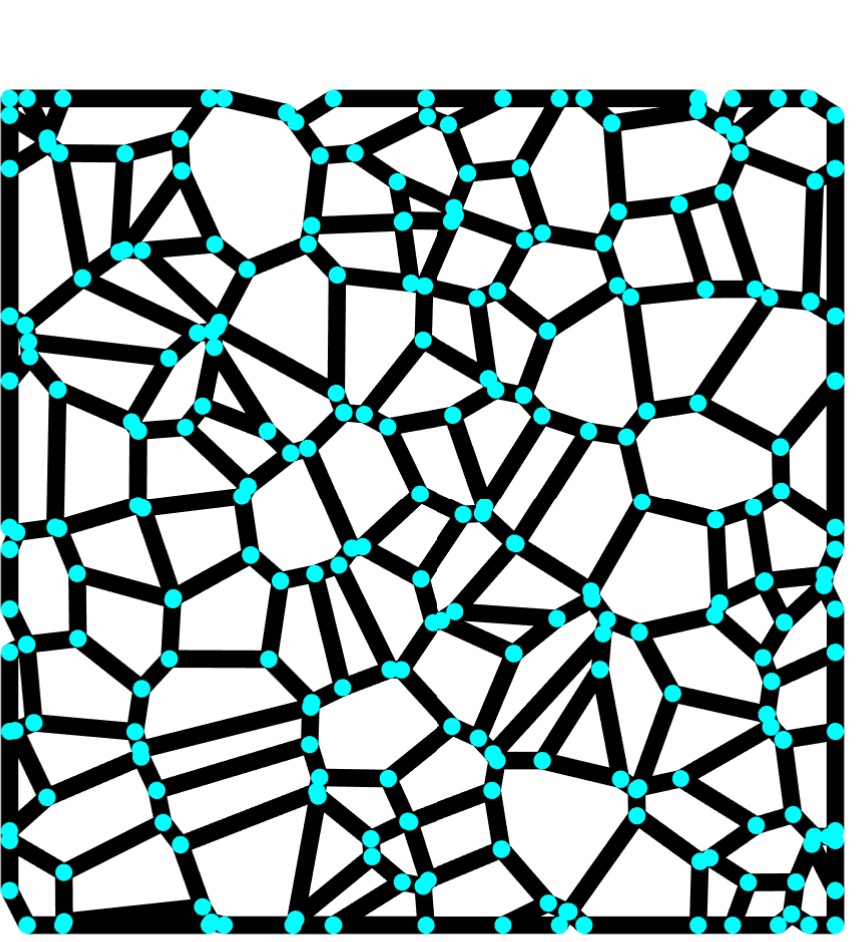}
    \caption{Lattices used in the analysis, arranged from left to right: triangular, square, and Voronoi lattices.}
    \label{fig:lattices}
\end{figure}

\begin{figure}[b]
    \centering
    \includegraphics[trim={0.5cm 0.5cm 1.5cm 1.5cm},clip,width=\linewidth]{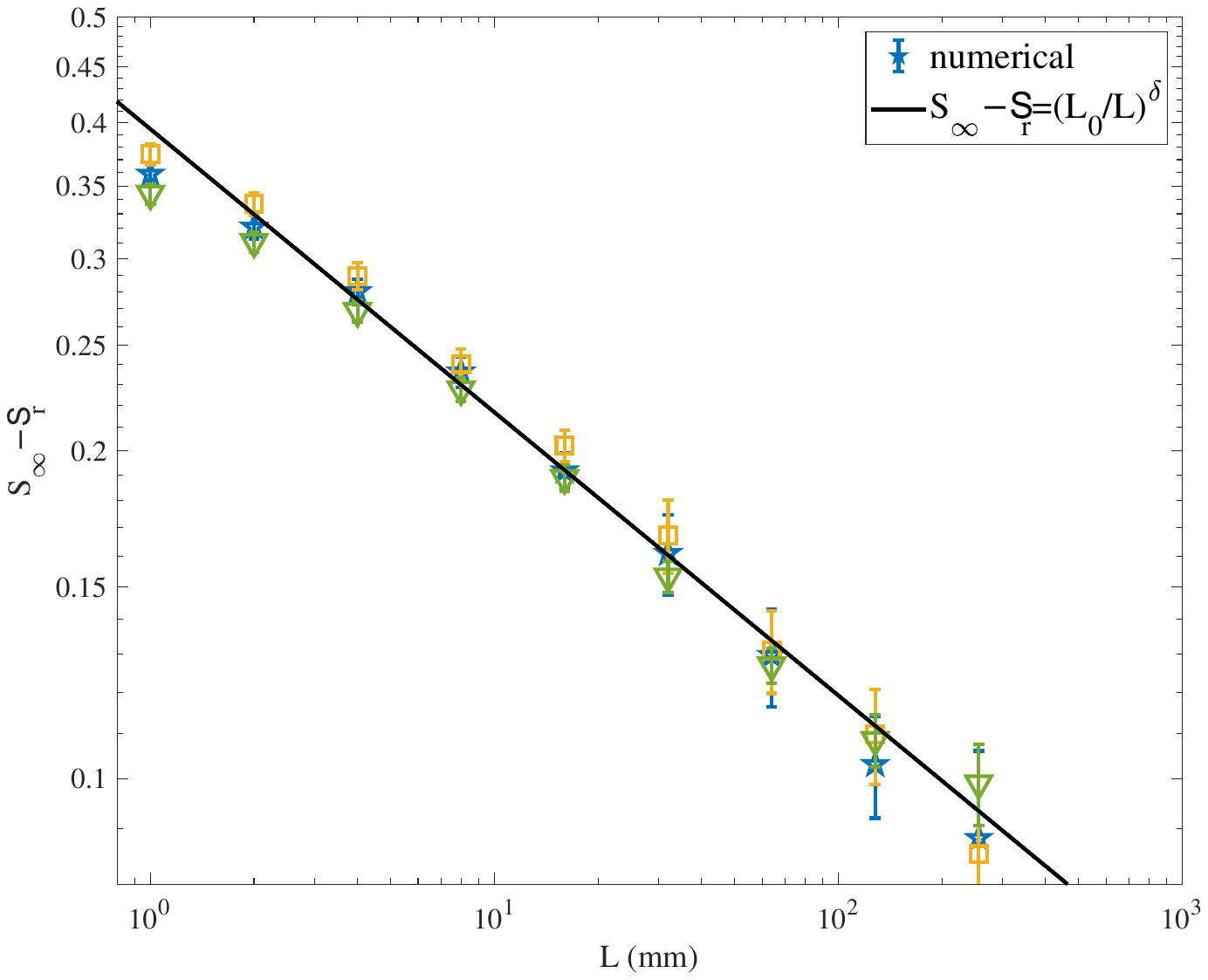}
    \caption{Residual saturation versus lattice size for triangular (triangles), square (squares), and Voronoi (stars) lattices.  $S_r(L)$ denotes the residual saturation for the lattice size $L$, while $S_\infty$ is the statistical limit of the residual saturation when $L\to\infty$. The fitted power law is given by Eq.~\ref{eq:residual}. }
    \label{fig:saturation}
\end{figure}

Similar finite-size effects as in BP are observed in our simulations. The percolation transition becomes sharper as the graph size increases, which is a characteristic of critical systems. In addition, strong statistical fluctuations in the residual saturation are observed from sample to sample, so a large number of simulations with the same parameters set are required to achieve averaged representative pressure-saturation curves. An extensive set of simulations was performed using three different networks: square, triangular, and Voronoi; see Figure~\ref{fig:lattices}. These lattices have a well-known BP threshold \cite{kesten1980critical,sykes1963some,becker2009percolation}, a well-defined coordination number, and belong to the same BP universality class; see Table \ref{tab:lattice_params}. For the scale analysis, we define the lattice size as $L=\sqrt{N_p}$, where $N_p$ is the number of pores. To examine the dependence of the pressure-saturation relations on the lattice size, we varied it as $L=10\times 2^n$, where $n=0,1,\dots8$. For each case, simulations were performed on $960$ random realizations. In each realization, the throat radii were generated using a log-normal distribution. After simulations, the averaged values and standard errors were calculated.

\begin{figure}[t]
    \centering
    \includegraphics[trim={0cm 4cm 0cm 5cm},clip,width = \linewidth]{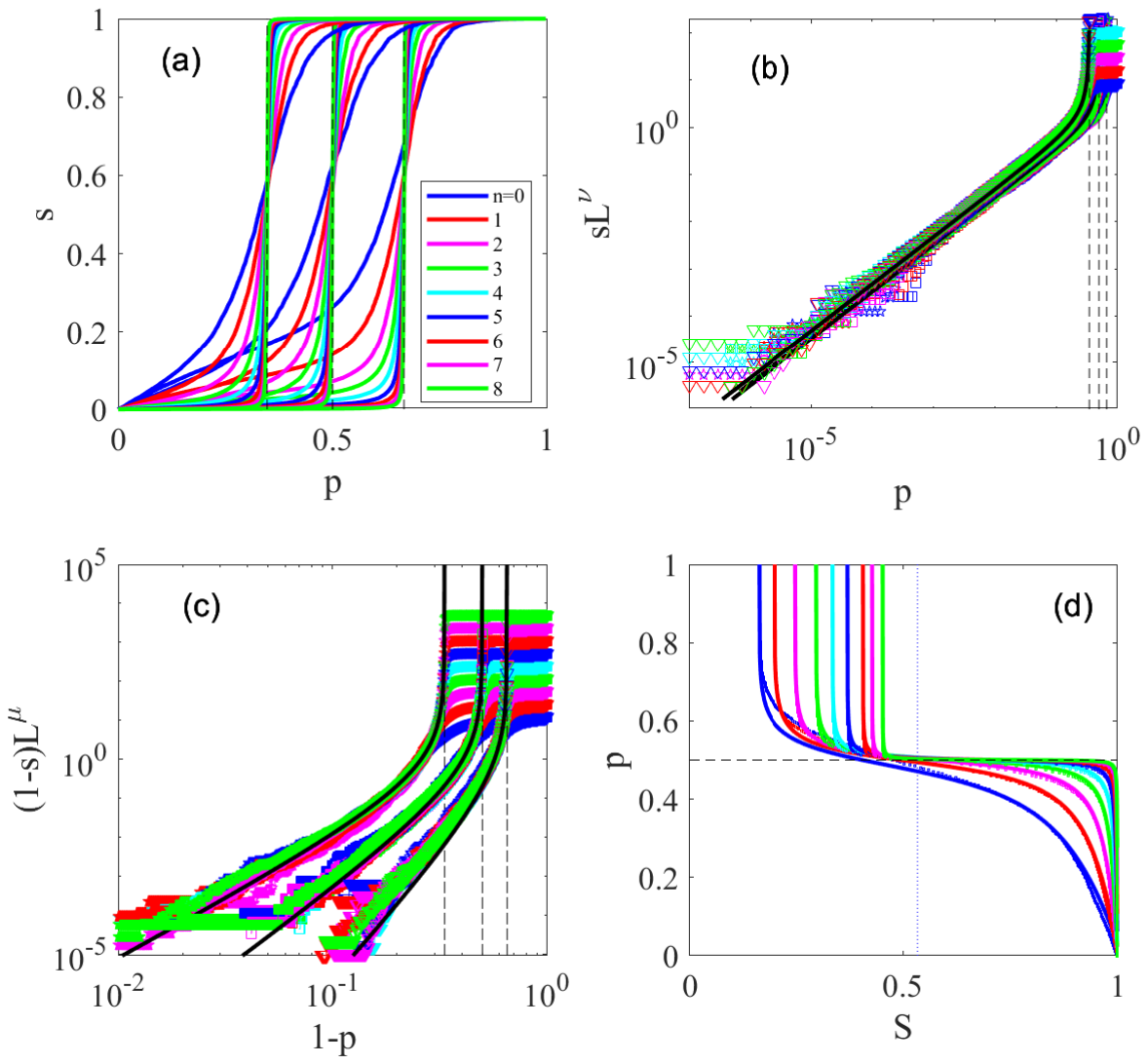}
    \caption{Results from the finite-size scaling analysis of percolation: (a) Normalized saturation $s$ versus occupancy probability $p$ for different lattice sizes $L = 10 \times 2^n$, where $n = 0, 1, \ldots, 8$. The lattices correspond to those shown in Fig.\ref{fig:lattices}. As $L$ increases, the percolation threshold manifests as a sharp phase transition at $p_c \approx 0.35$, 0.5, and 0.66 for triangular, square, and Voronoi lattices, respectively. (b) $s L^\nu$ versus $p$ for all values of $L$, demonstrating data collapse across all lattice sizes for triangular (triangles), square (squares), and Voronoi (stars) lattices. (c) $(1 - s) L^\mu$ versus $1 - p$ for all lattice sizes, with data collapse depending on the coordination number of each lattice. (d) The uniform saturation $S = 1 - s(1 - S_r)$ calculated from the numerical data (symbols) and the analytical solution of Eq. \ref{eq:uniform} (solid lines) of square lattices. The dashed lines indicate the classical bond percolation (BP) threshold. The dotted line represents the residual saturation for infinity systems.}
    \label{fig:mosaic}
\end{figure}

The relation between residual saturation and lattice size is shown in Fig.~\ref{fig:saturation}. For all cases, the same power-law exponent $\delta = 0.26\pm0.01$ is obtained, suggesting it as a universal exponent within the universality class of regular two-dimensional networks.
The exponent is significantly small, indicating strong size-dependency effects and a slow convergence to the statistical limit. The best fit for the residual saturation is given by 
\begin{equation}
    S_r(L)  \approx S_\infty - \left(\frac{L_0}{L}\right)^\delta,
    \label{eq:residual}
\end{equation}
where $S_\infty = 0.53 \pm 0.04$ is the statistical limit of the residual saturation of the resident fluid, and $L_0=0.25\pm0.1$. This power-law relation is inherent in the pore morphology method and differs from that obtained in IP \cite{knackstedt2001invasion}. In short, the power-law relation in two-dimensional lattices may be topologically invariant and universal in pressure-controlled cases but not in flow-rate-controlled experiments.

A critical issue in the simulation of two-dimensional lattices is the slow convergence of the residual saturation to the statistical limit. This has implications on the finite-size effects on the representative element volumes (REV) used to obtain the constitutive relations for large-scale simulations. From Eq.~\ref{eq:residual}, the deviation of the residual saturation from its statistical limit is given by $\Delta S \sim L^{-0.26}$. This means that to reduce this deviation by 50\%, the REV size must be increased 14 times! This poses a serious question about the way we perform large-scale simulations. These simulations are based on the assumption that the constitutive relations at each grid point are obtained from REVs that are large enough to avoid size effects. However, these effects can only be removed by REV sizes larger than the grid size required to obtain accurate results, establishing a fundamental limit in the accuracy of these simulations. We note that the power-law exponent in IP depends on the dimensionality of the sample, and the convergence to the statistical limit in IP is faster for higher-dimensional lattices \cite{dias1986percolation}. New light on this fundamental issue can be gained by the pore morphology analysis of other networks, such as 3D, space-correlated, multiscale, and small-world networks. 

\subsection{Self-similar solutions}

Self-similarity is a remarkable property of certain systems looking statistically the same on different scales. In percolation theory, this is usually demonstrated by scaling to finite sizes \cite{stauffer2018introduction,sampaio2018elastic}. After the order parameter is defined, data from different system sizes can be rescaled so that they collapse onto a single universal curve. This collapse shows that the order parameter evolves in the same way near the percolation threshold, regardless of the microscopic details or the absolute size of the system, confirming the universal character of the transition. In our percolation model, we use the normalized saturation as an order parameter. Unlike the typical self-similarity analysis of percolation, we demonstrate different self-similarity rules in the sub-/supercritical regimes, and we match the self-similar solution using asymptotic matching.

Based on residual saturation, normalized saturation $s(p,L)$ is defined as the order parameter of the percolation process. This is expressed in terms of the lattice site $L$ and the occupancy probability $p$ 
\begin{equation}
    s(p,L)=\frac{1-S(p,L)}{1-S_r(L)},
    \label{eq:saturation}
\end{equation}
where $S(p,L)$ is the saturation of the resident fluid as a function of the occupancy probability $p$ and $S_r(L)$ is its residual saturation. In percolation theory, the standard procedure involves performing a finite-size scaling analysis of the relation between the order parameter $s$ and the control parameter $p$ for values close to the critical percolation $p_c$ where a behavior $s \sim |p-p_c|^{-\beta}$ is expected as $p\to p_c$ \cite{stauffer2018introduction}. For our simulations, the $s$-$p$ curves for different lattice sizes are shown in Fig.~\ref{fig:mosaic}a. The curves suggest self-similar solutions in the subcritical ($p<p_c$) and supercritical ($p>p_c$) regimes, with a clear asymmetry in both solutions. To find the self-similar solutions, an asymptotic matching analysis is proposed: first, by finding the scaling relations in each regime, and then by asymptotically matching the solutions. 
For the inner or subcritical solution, ($p<p_c$), the following scaling law is proposed.
\begin{equation}
    s = \frac{f(p)}{L^\nu(p_c-p)^\alpha},\qquad p<p_c
    \label{eq:inner}
\end{equation}
The outer, or supercritical solution ($p>p_c$), is expected to satisfy the following.
\begin{equation}
    1-s = \frac{g(1-p)}{L^\mu(p-p_c)^\beta},,\qquad p>p_c
    \label{eq:outer}
\end{equation}
Figs.~\ref{fig:mosaic}(b-c) show a very good collapse of the data in the sub-/supercritical regimes. The exponents are determined as follows. The exponents $\nu$ and $\mu$ are optimized to achieve the best collapse of the curves $s/L^\nu$ and $(1-s)/L^\mu$ in the respective regimes. Then the exponents $\alpha$ and $\beta$ are the minimal positive values that eliminate the singularity of the self-similar functions $f(p)$ and $g(1-p)$ at $p=p_c$. The calculated exponents are $\nu = 0.92\pm 0.02$, $\mu = 1.08\pm0.2$, $\alpha=1.13\pm0.02$, and $\beta=1.35\pm0.02$. These exponents are consistent across all three lattice topologies examined in this study, suggesting that they are universal within the universality class of regular two-dimensional lattices.

 The self-similar scaling functions $f(p)$ and $g(1-p)$, on the other hand, are lattice-dependent and should be estimated from the collapsed curves. 
 It is observed that $f(p)\sim p$ as $p \to 0^+$ and $g(1-p)\sim (1-p)^z$ as $p \to 1^-$, where $z$ is the coordination number of the lattice. The self-similar function $f(p)$ is approximated as the Taylor expansion around $p=0$ and the function $g(1-p)$ is approximated using the power expansion of 
 $(1-p)^{z}$ around $p=1$. The exponents of Taylor expansions are calculated using the pressure-saturation relation of the largest lattice, which has size $L=10\times 2^8=2560$. The results shown in Table~\ref{tab:lattice_params}, suggest that $g(1-p)$ is clearly non-universal since it depends on the coordination number of the lattice. The collapse of the curves into $f(p)$ in Figure~\ref{fig:mosaic}b signals universality, however, whether $f(p)$ itself is universal remains inconclusive based on the data presented in Table~\ref{tab:lattice_params}.

\subsection{Asymptotic matching}
Asymptotic matching is a technique used in boundary layer theory that connects solutions that are valid in different regions by requiring agreement on their overlap \cite{arias2018closed}. An {\it inner solution} captures behavior near a boundary, while an {\it outer solution} describes behavior far away. By ensuring that they agree in the overlap, a single uniform solution valid across the domain is obtained. This approach will be applied to the percolation problem to match the inner and outer scaling regimes into a uniform solution.

The first step for asymptotic matching is the identification of the boundary layers. Eq.~\ref{eq:inner} defines $p$ implicitly in terms of $s$ as $p = p_{inner}(s)$. In the same way, $p$ is implicitly given in terms of $s$ from Eq.~\ref{eq:outer}, which is expressed as $p = p_{outer}(s)$. The uniform solution is then obtained by matching both solutions.
\begin{equation}
    p_{unif}(s) = p_{inner}(s)+p_{outer}(s)-p_c.
    \label{eq:uniform}
\end{equation}
The matching solution is compared with the numerical data in Fig.~\ref{fig:mosaic}(d). An excellent agreement between the numerical results and the self-similar closed-form solution is achieved. These solutions assume the classical bond percolation threshold $p_c$ as the overlap for the matching solution. Note that this matching does not account for the backbone percolation transition discussed in the literature. For example, in the backbone percolation model, a backbone phase transition is encountered above the classical percolation threshold \protect\cite{sampaio2018elastic}. If there is a backbone-like transition in two-dimensional lattices, this should be too close to the classical percolation transition to be detected by using our analysis. For practical purposes, both the percolation and the backbone transitions occur at the same value of $p_c$, and are reflected in the plateau of the pressure saturation relation in Figure \ref{fig:mosaic} d.  

\begin{table*}
    \centering
    \begin{tabular}{lccccc}
        \toprule
        Lattice    & $p_c$  & z & a1     & a2   & b       \\
        \midrule
        Voronoi    & $0.666$ & 3 & $2.3$ & $-3.2$ &  $1.7$ \\
        Square     & $0.5$   & 4 & $1.5$ & $-2.1$ &  $1.5$ \\
        Triangular & $0.3473$& 6 & $1.4$ & $-2.6$ & $1.0$  \\
        \bottomrule
    \end{tabular}
    \caption{ Bond percolation threshold $p_c$, coordination number $z$, and relevant coefficient of the Taylor expansion of the self-similar functions of the pressure-saturation relationships for different lattices. The expansions are given by $f(x)\approx a_1x+a_2x^2$ and $g(x) \approx bx^z$.
    The dependency of these functions on coordination number indicates that they are non-universal.}
    \label{tab:lattice_params}
\end{table*}

\section{Conclusions}
\label{sec:conclusion}
In summary, the pore morphology method is formulated on the undirected graph of the pore network, with normalized saturation serving as the order parameter. The connectivity of nodes and edges dictates the critical transitions, making the method an explicit application of topology. Two key topological invariants emerge from this analysis: the percolation threshold and residual saturation. The pressure–saturation relationship can be derived from scaling arguments, expressed in terms of scaling exponents. These exponents are universal, while the self-similar functions that describe the supercritical state are nonuniversal, depending on the coordination number of the lattice.

The significance of the proposed percolation model and the numerical results lies in the combination of topological invariance and universality. Because the critical exponents are independent of geometry and material properties, they enable the formulation of general laws for multiphase flow that are widely applicable. At the same time, the explicit expression for the occupation probability as a function of capillary pressure, pore-size distribution, and interfacial parameters provides an analytical link to measurable material properties.

Viewed through the lens of topology, the percolation threshold can be understood as a topological phase transition. Below the threshold, the pore network for infinite systems essentially consists of two connected clusters of defending and invading fluids. At the threshold, a giant connected component for the invading phase emerges, while the defending fluid breaks into many disconnected clusters. This interpretation shows that the percolation transition is not merely probabilistic or geometric but a fundamental change in the topology of the network.
The universality of scaling exponents across lattices of different geometry further supports the existence of topological universality classes, in which systems with equivalent connectivity exhibit identical critical behavior. This universality highlights that topology, not geometry, governs the essential physics of the invasion process.

The invariants identified here are significant because they remain unchanged under variations in geometry or material properties. The percolation threshold and residual saturation thus offer reliable descriptors of pore-scale invasion, and their persistence makes them a cornerstone for developing upscaling theories of multiphase flow that require generality and independence from microscopic details.

\section{Acknowledgments}
The author thanks Morteza N. Najafi for useful discussions.




\bibliography{sn-bibliography}
\end{document}